\documentstyle[12pt]{article}

\input{epsf.sty}
\def\framegraphics{\def\ifframe{\iftrue}}
\def\dontframegraphics{\def\ifframe{\iffalse}}
\def\drawgraphics{\def\ifdraw{\iftrue}}
\def\dontdrawgraphics{\def\ifdraw{\iffalse}}

\dontframegraphics
\drawgraphics

\newcommand{\graphics}[6]{
\def\epsfsize##1##2{#6##1}
\begin{picture}(#2,#3)
  \ifframe
    \put(0,0){\framebox(#2,#3){}}
  \fi
  \ifdraw
    \put(0,#3){\begin{picture}(0,0)
                 \put(#4,#5){\epsfbox{#1}}
               \end{picture}}
  \fi
\end{picture}}

\begin{document}

ADP-94-27/T166  \hspace*{6cm} December 1994

\vspace{2cm}
\centerline{\large A Study of Structure Functions for the Bag
Beyond Leading Order}
\vspace{.4cm}
\centerline{F.M. Steffens and A.W. Thomas}
\vspace{1cm}
\centerline{Department of Physics and Mathematical Physics}
\centerline{University of Adelaide}
\centerline{Adelaide, S.A. 5005, Australia}
\vspace{.3cm}
\begin{abstract}
There has recently been surprising progress
in understanding the spin and flavor dependence of deep
inelastic structure functions in terms of the same physics
needed in the simple quark models used for hadronic
spectroscopy. However, the corresponding scale is usually
very low, casting doubt on the use of leading order $QCD$
evolution. We show that the conclusions are not significantly altered
if one goes to next-to-leading order. In particular, the excellent
agreement with unpolarized and polarized valence quark
distributions is retained.
\end{abstract}
\newpage

\section{Introduction}


Deep inelastic scattering continues to provide a wealth of surprising
and challenging information concerning the structure of nucleons and nuclei.
The nuclear EMC effect, now more than a decade old \cite{emc83}, provided
a dramatic challenge to our conventional view of nuclear structure
\cite{reviews}.
In particular, it suggested a systematic change in the valence quark structure
of a nucleon inside matter. This clearly goes to the heart of our understanting
of nuclear structure. More recently the European Muon Collaboration revealed
\cite{emc88} an unexpected deviation from the Ellis-Jaffe sum rule
\cite{ellis},
which became known as the ``spin crisis". Finally the New Muon Collaboration
\cite{nmc} has confirmed suggestions \cite{feynman}
of a violation of SU(2) flavour symmetry in the nucleon sea.

All of these observations demand theoretical interpretation. In the absence of
reliable
calculations using lattice QCD, and with QCD sum rules yielding only a
few moments, one is naturally led to the problem of relating low energy
quark models to the parton distributions measured in deep-inelastic
scattering. Given that deep inelastic data is valence dominated at low-$Q^2$,
it was suggested fairly early \cite{parisi} that the natural connection
was to use the quark model to calculate the leading twist parton
distributions at some low scale, $\mu^2$, and to evolve them to a higher
scale, $Q^2$, using the renormalization group equations (RGE) and
then compare the resulting distributions with data \cite{parisi}.
In the context of the bag model,
Jaffe \cite{jaffe1} and Le Yaouanc et al. \cite{yaouanc} were the first
to implement this idea, followed
by Hughes \cite{hughes} and Bell \cite{bell} and others \cite{benesh}.

A major problem in these calculations was the lack of correct
support for the parton distributions.( A quark distribution
has to vanish outside the region $0\leq x \leq 1$.) This disease was
cured some time ago in a series of papers by Signal, Thomas and
Schreiber \cite{signal,andreas}. The proposal was to construct the
parton distributions using the formal analysis of Jaffe \cite{jaffe2},
and to ensure energy-momentum conservation before any approximations were made.
(To better appreciate the importance of correct support in the distributions,
the reader is referred to ref. \cite{fernando}.)
In ref. \cite{andreas} a detailed study of the MIT quark
distributions was made. Various distributions, polarized and unpolarized, were
calculated at the model scale and then evolved in leading order (LO),
using the RGE, to the scale of the experiment. A very good, qualitative
agreement was obtained, which was rewarding and stimulating given the
simplicity of the model. It became clear that high energy experiments could
be, to a large extent, explained using as input low energy physics.

Of course, bag model calculations using the method of ref. \cite{andreas}
are not the only ones that can be
found in the literature (see, for example, ref. \cite{song}). Moreover, there
are
calculations based on other models. For example, there are works on the
colour-dieletric model \cite {barone} or on the
non-relativistic quark model \cite{fernando,nrq}. However, besides the fact
that some of
the calculations suffer from the problem
of poor support, it is known \cite{fernando} that the parton distributions
constructed from Gaussian wave functions
do not have the right behavior as $x$ approaches 1.
Recently, there have also been attempts to formulate the problem in terms of
relativistic quark-nucleon vertex functions \cite{mulders,wally}.

Thus far, all analyses of the $x$ dependence of parton distributions
within quark models have been performed only in leading order. There are
numerous
studies of higher order evolution of unpolarized data in the form of
parametrizations \cite{gluck,mrs}, or higher order corrections to
the first moment
of some QCD sum rules \cite{find}, but no study of models involving
both polarized and unpolarized data \footnote{In ref.\cite{neerven}
the $x$ dependence of $g_{1p}(x)$ was calculated in NLO.
However, parametrizations of the data were used for the input}.
Besides the fact that  corrections
of higher order in the strong coupling constant $\alpha_s(Q^2)$ are
already important per se, we emphasise that
a next-to-leading order (NLO) analysis is necessary because
the scale of the quark model is generally very low so that the
applicability of the leading order RGE is questionable.
In this paper, we will apply the NLO evolution to the polarized and
unpolarized valence distributions in the proton and to the $x$ dependence
of the Bjorken sum rule.
We will compare the resulting distribution with the MRS \cite{mrs}
parametrizations and with preliminary data for the
polarization of valence quarks in the proton \cite{wislicki}.
The paper is organized as follows. In Sec. II we briefly review
the process of QCD evolution in NLO. In Sec. III we review the procedure for
calculating quark distributions in the bag model. In Sec. IV we
present some numerical results, while Sec. V is used to present our
conclusions.

\section{Distributions at Next to Leading Order}

It is well known that one can write
the moments of the structure functions as:

\begin{equation}
M_n(Q^2)=\sum_i C_n^i (Q^2/\mu^2, g) A_n^i (\mu^2).
\label{II1}
\end{equation}
The sum runs over all spin-n, twist-2 operators. In this work
we will restrict ourselves to the nonsinglet sector because
the complete, singlet anomalous dimensions in two loops for polarized
scattering are not known.
(Up to now, only the quark-quark and the quark-gluon part of the singlet
Altarelli-Parisi splitting functions have been calculated \cite{neerven}.)

Equation (\ref{II1}) is a direct result of the operator product expansion
applied to the forward scattering of the photon from the hadronic
state through the e.m. current, $J_\mu$.
The amplitudes ${\it A}_n^i (\mu^2)$ are target dependent and involve
non-perturbative QCD. They are related to the moments of the
quark distribution in the target at the renormalization scale $\mu^2$.
The Wilson coefficients, $C_n^i (Q^2/\mu^2, g)$, are target independent
and calculable in perturbation theory. They carry all the information
about the scale dependence of the structure functions and are
well determined. The
evolution in $Q^2$ for the Wilson coefficients is given by the solution
of the corresponding renormalization group equation which has the form:

\begin{equation}
C_n (Q^2/\mu^2, g) = C_n (1,
\overline{g}^2)exp\left(-\int_{\overline{g}(\mu^2)}^
{\overline{g}(Q^2)}
dg' \frac{\gamma^n (g')}{\beta(g')}\right).
\label{II2}
\end{equation}
Here, $\overline{g}(Q^2)$ is the running coupling constant (such that
$\overline{g}(Q^2=\mu^2)=g$), $\gamma^n (g)$ is the anomalous dimension
of the corresponding nonsinglet operator and $\beta(g)$ is the QCD beta
function.
Expanding to second order they are expressed as:

\begin{equation}
\beta(g)=-\beta_0 \frac{g^3}{16\pi^2} - \beta_1 \frac{g^5}{(16 \pi^2)^2},
\label{II3}
\end{equation}
\begin{equation}
\gamma^n (Q^2)=\gamma^{(0) n} \frac{g^2}{16\pi^2} + \gamma^{(1) n}
\frac{g^4}{(16\pi^2)^2},
\label{II4}
\end{equation}
\begin{equation}
C^n = 1+F^n  \frac{g^2}{16\pi^2},
\label{II5}
\end{equation}
while the running coupling constant is determined by solving the
transcendental equation:
\begin{equation}
ln\frac{Q^2}{\Lambda^2}=\frac{16\pi^2}{\beta_0 \overline{g}^2} -
\frac{\beta_1}{\beta_0^2} ln\left[\frac{16\pi^2}{\beta_0 \overline{g}^2} +
\frac{\beta_1}{\beta_0^2}\right].
\label{II6}
\end{equation}

The QCD scale parameter, $\Lambda$, is determined by comparing
the theoretical calculations with the experimental data, and $\beta_0 ,\;
\beta_1 ,\; \gamma^{(0) n} , \; \gamma^{(1) n}$ and $F^n$ are parameters
determined in perturbation theory. Their form is listed in the appendix.
Throughout this paper we will use $\Lambda=0.2\;   GeV$ (which is
within the experimental errors \cite{particle}). For unpolarized
scattering both the leading order (LO) \cite{gross} and next to leading
order (NLO) \cite{jones} coefficients are well known.
One need only be careful with the renormalization scheme dependence:
although $\beta_0$, $\beta_1$ and $\gamma^{(0)}$ do not depend on the scheme,
$\gamma^{(1) n}$ and $F^n$ do, and both have to be calculated in the
same scheme in order to obtain physically meaningful results \cite{gluck2}.
For the nonsinglet operators, all coefficients but $F^n$ are the
same for both polarized and unpolarized scattering. The Wilson
coefficient $F_n$ was first calculated in the $\overline{MS}$ scheme
by J. Kodaira and collaborators \cite{kodaira}. With this in mind we
write the NLO evolution equation for the moments with the help of expressions
(\ref{II2})-(\ref{II5}):

\begin{eqnarray}
M_n (Q^2)&=&A_n(\mu^2)\left[\frac{\overline{g}^2(Q^2)}
{\overline{g}^2(\mu^2)}\right]^{\gamma^{(0) n}/2\beta_0} \nonumber \\*
         &\space&\left(1+\frac{\overline{g}^2(Q^2)}{16\pi^2} F^n +
\frac{(\overline{g}^2(Q^2)-\overline{g}^2(\mu^2))}{16\pi^2}
\left(\frac{\gamma^{(1) n}}{2\beta_0}-\frac{\beta_1 \gamma^{(0)
n}}{2\beta_0^2}\right)
\right).
\label{II7}
\end{eqnarray}
This equation can be rewritten as:

\begin{equation}
M_n(Q^2)=A_n(Q^2)C_n(1,\overline{g}(Q^2)),
\label{II8}
\end{equation}
which makes it clear that the moments of the quark distributions in NLO have
the same $Q^2$ dependence as in LO, but that this is not the same for
the moments of the structure functions. In other words, $A_n(Q^2)$ gives the
moments of the quark distributions and $M_n(Q^2)$ gives the moments of the
structure
functions. Expression (\ref{II8}) also expresses clearly the scheme dependence
of the quark distributions beyond leading order, which means that they are
unphysical.
{}From eq. (\ref{II7}) is
straightforward to verify that in NLO, QCD sum rules, like the Bjorken sum
rule,
pick up $\overline{g}^2$ corrections.

\section{The Quark Distributions for the Bag Model}

The matrix elements ${\it A}_n(\mu^2)$ can be written in terms of parton
distributions,
as showed by Jaffe \cite{jaffe}. This is done by defining the distribution
as the integral,
in the light cone gauge, of the forward virtual quark-target scattering
amplitude over all the parton momenta (in the light cone) but keeping the
plus component fixed and equal to $x p^+$ (with $x$ the Bjorken variable and
$p^+$ the plus component of the target momentum). The parton distribution
written in this way has support only for $-1\leq x \leq 1$. Explicitly,

\begin{equation}
q(x)=p^+ \sum_n \delta (p^+ (1-x)-p_{n}^{+})\mid\langle n| \Psi_+ (0)|
     p\rangle\mid ^2,
\label{II9}
\end{equation}
and

\begin{equation}
{\it A}_{n}=\int_{-\infty}^{+\infty} q(x)x^{n-1}dx,
\label{II10}
\end{equation}
where $p_n$ is the momentum carried by the spectators to the
interaction (intermediate states). An equation very similar to equation
(\ref{II9})
can be written for the antiquark distribution, but with $\Psi^{\dagger}$
replacing $\Psi$. There are two kind of
contributions to (\ref{II10}): one coming from the annihilation of
one of the quarks in the nucleon and the other from the creation
of a quark or antiquark in the nucleon. The first process is what we call
the two quark contribution and it is the dominant one. We show this
contribution within the model, in question, in detail below (see equ.
(\ref{II11})).
The second contribution is a four quark one. As will become clear
below, this kind of contribution will be used to fix the normalization
problem of the quark distributions in the bag but it will not be calculated
explicitly
within the model.

The approach followed in this paper to calculate $q(x)$ will be that developed
by the group at the University of Adelaide \cite{signal,andreas}, in which
the MIT \cite{mit} bag model wave functions are used and the correct
support for the distribution is fully assured. Basically, what is done is to
project the moment vectors $|p\rangle$ in the coordinate space and
calculate these overlaps using the Peierls-Yoccoz aproach \cite{peierls}.
This ensures that the wave functions are momentum eigenstates, so
attacking one of the major problems of the bag model, that is the lack of
translational invariance. The Peierls-Yoccoz projection is regulated by
the weight functions $\phi(\vec p)$ given by:

\begin{equation}
\mid\phi_l (\vec p )\mid^2 = \int d\vec x e^{-i\vec p\cdot\vec x}\left[\int
d\vec y
           \psi^{\dagger}(\vec y - \vec x)\psi (\vec y)\right]^l .
\label{II10a}
\end{equation}
Inserting a complete set of position states, converting the sum into
an integral over the momentum of the intermediate states and using the
projection
given above, we get for the two quark piece (full details can be
found in ref. \cite{andreas}):

\begin{eqnarray}
q^{\uparrow \downarrow}_f (x)&=&\frac{M}{(2\pi)^2}\sum_m
\langle\mu|P_{f,m}|\mu\rangle
\nonumber \\*
     &\times&\int_{[M^2 (1-x)^2 -M_{n}^{2}]/2M(1-x)}^{+\infty}\mid\vec{p}_n\mid
     d\mid\vec{p}_n\mid \frac{|\phi_2(\vec{p}_n)|^2}{|\phi_3(0)|^2}|
     \tilde\psi^{\uparrow \downarrow}_{m}(\vec{p}_n)|^2 .
\label{II11}
\end{eqnarray}
Here $|\mu\rangle$ is the spin-flavor part of the wave function of the initial
state (at rest), $P_{f,m}$ makes the projection onto flavor $f$ and spin
projection
$m$, $M_n$ is the mass of the intermediate state and $\tilde\psi$ the Fourier
transform of
$\psi$. The expression for the four quark contribution has the same form, but
$\phi_2$ is replaced by $\phi_4$.

Once we have a model to calculate the matrix elements $A_n$ we can proceed
to evalute the moments of the structure functions. Before doing so,
a few remarks need to be made. The first deals with the spin
of the intermediate states. In the case of a two-quark spectator system,
the sum of the spins may be zero (a scalar diquark with a lower mass $M_s$), or
one
(a vector diquark with a higher mass $M_v$). The mass difference between these
states is found to be around $200 MeV$ in most quark models, and this is
crucial to understanding the soin and flavour dependence of the distributions
\cite{close}.
Taking this into account,
one can calculate the contributions to the quark distribution (\ref{II11}) due
to the scalar and vector diquark systems, by using the $SU(6)$ wave function
for
$|\mu\rangle$. The second issue is the form of
the four quark contibution. It is known \cite{andreas} that the bag model has
problems to fix the normalization of the quark distributions. Probably this is
due the fact that we do not calculate the masses of the intermediate states
in a way consistent with the model. To avoid this problem, we choose the four
quark
term to have the form $(1-x)^7$ and normalize it in a way that the overall
quark distribution satisfies the normalization condition. Although this
procedure is clearly phenomenological and not directly related to the model,
we note that a term of this form will affect only the region $x<0.3$.
Furthermore its shape resembles very much
the actual form of the the four quark term calculated using the same
formalism as that used for the two quark contribution.

\section{Results}

Now that we know how to calculate the matrix elements $A_n$ we
can proceed to apply the evolution equations to this particular model.
The procedure we will follow is: first, we find a best fit for the
unpolarized valence distribution by comparing the LO and NLO evolution
against the MRS \cite{mrs} parametrization of the experimental data
at $10\; GeV^2$. In doing that we fix the set of model parameters, e.g., radius
of
the bag and masses of the intermediate states. Using the virial theorem
in the bag, where the bag itself carries the same energy as each
of the three quarks confined in the bag, we roughly expect $M_s \simeq 750\;
MeV
- 150\; MeV$ and $M_v \simeq 750\; MeV + 50\; MeV$. But this is, of course,
only
an approximation and we have some freedom around these values.
We  also
determine the scale $\mu$ at which the model is supposed to be
valid. Using this set of parameters, we can then calculate the predictions
of the model for other quantities. Here, as we have only
the nonsinglet evolution for NLO, we will restrict ourselves to the polarized
valence distribution of the proton and to the difference between
the spin structure functions of the proton and neutron.

\begin{figure}[h]
\vspace{5cm}
\graphics{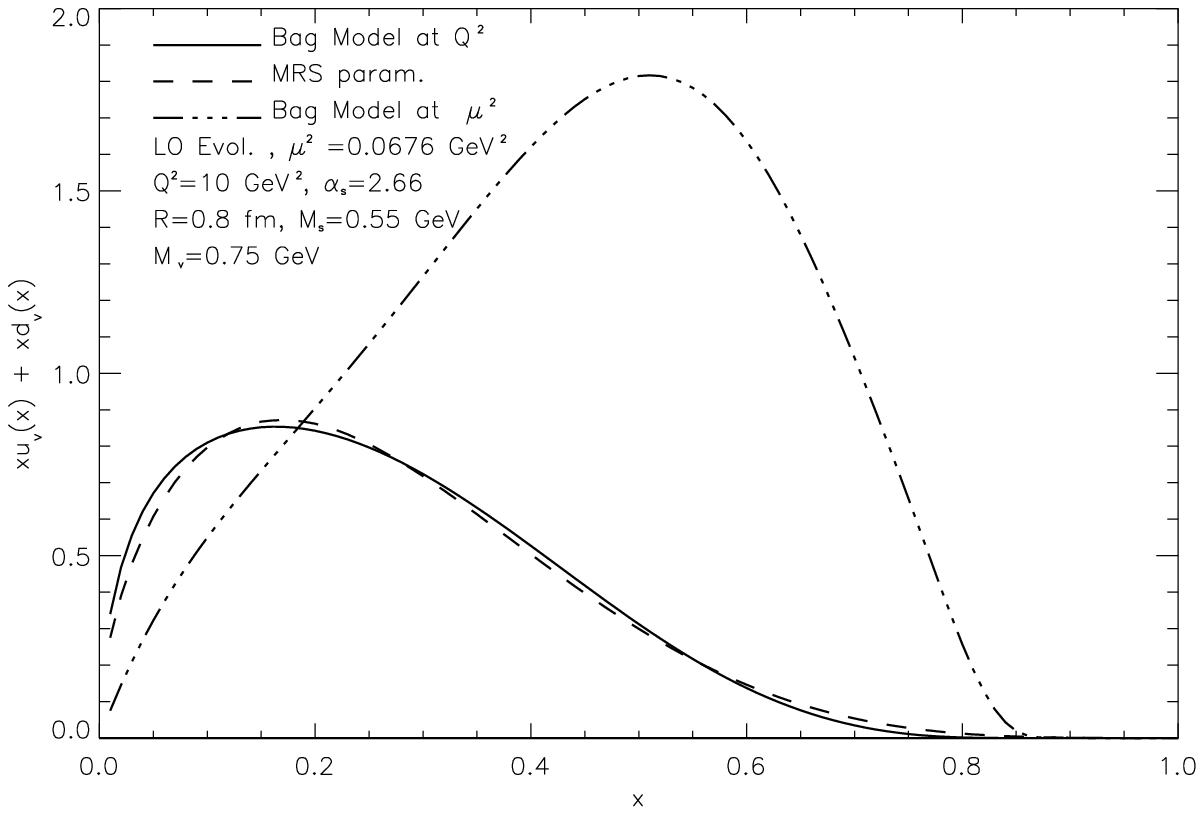}{17}{10}{0}{-10}{1}
\caption{}
\label{fig1}
\end{figure}

\begin{figure}[h]
\vspace{5cm}
\graphics{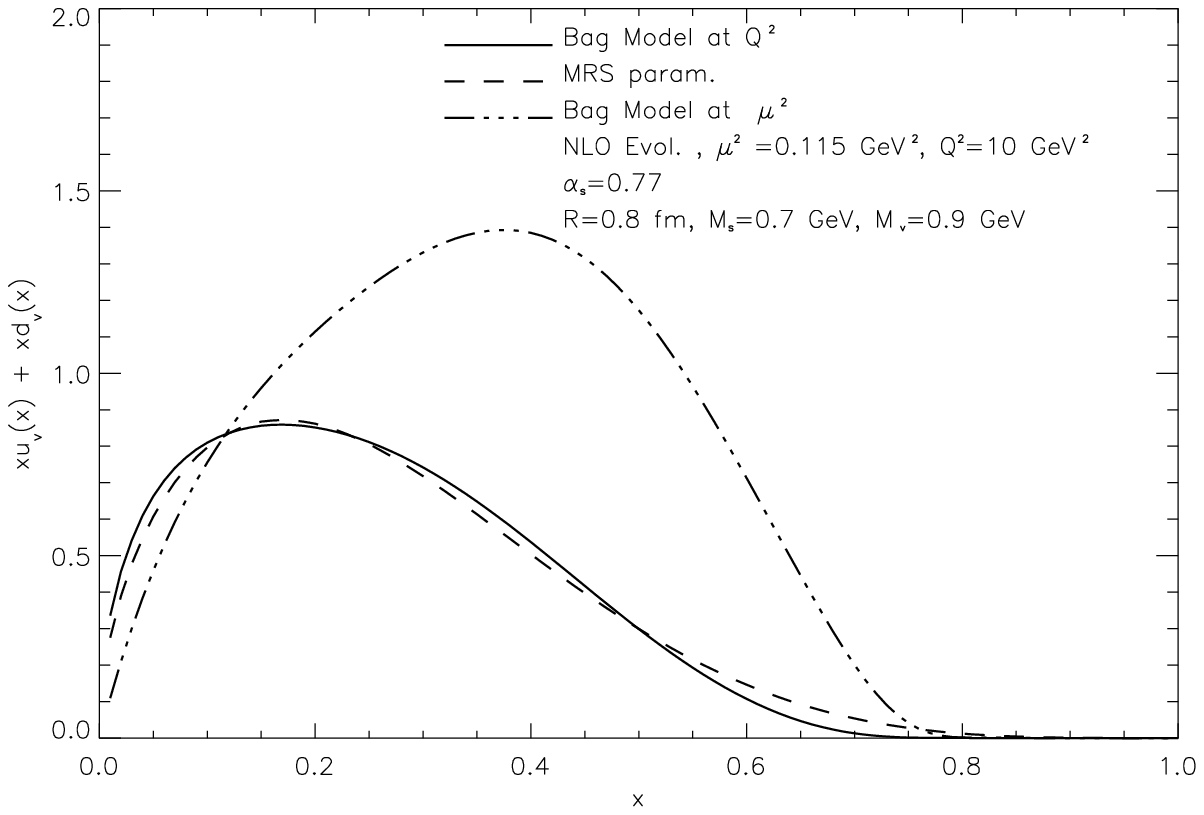}{17}{10}{0}{-10}{1}
\caption{}
\label{fig2}
\end{figure}

In LO it is found that for $R=0.8\; fm$, $M_s =0.55\; GeV$, $M_v =0.75\; GeV$
and
$\mu^2 =0.0676\; GeV^2$, the unpolarized valence distribution of the proton in
the bag
model fits the MRS parametrization of the data. The result is displayed in Fig.
\ref{fig1} together with the bag model distribution at scale $\mu^2$. The
agreement
is excellent with no significant discrepancy. However, at such a low scale
the coupling constant is $\alpha(Q^2=\mu^2)=2.66$. This is a rather
large value (as found in ref. \cite{andreas}), and raises doubts about the
whole
procedure.
In NLO the parameters  used to fit the unpolarized
valence distribution have the following values: $R=0.8\; fm$, $M_s =0.7\; GeV$,
$M_v =0.9\; GeV$ and $\mu^2 =0.115\; GeV^2$. We notice that the
radius for LO and NLO is the same but not the masses, being slightly bigger in
the
latter case. This is because in NLO one need not evolve as far, the
scale $\mu^2$ is larger and so the masses of the intermediate states
need to be larger (the parton distributions peak at a larger $x$ for
smaller intermediate state masses). In Fig. \ref{fig2} the
NLO results are shown and again we see a very good agreement between the
calculated valence distribution and the MRS parametrization of the data.
What is remarkable now is the fact that the strong coupling  constant
drops to $0.77$ at $\mu^2 =0.115$ in NLO. Although this value is still
large, it is much smaller than the LO case, making the evolution more
reliable. Both LO and NLO moments are evolved to $Q^2=10\; GeV^2$.

\begin{figure}[h]
\vspace{5cm}
\graphics{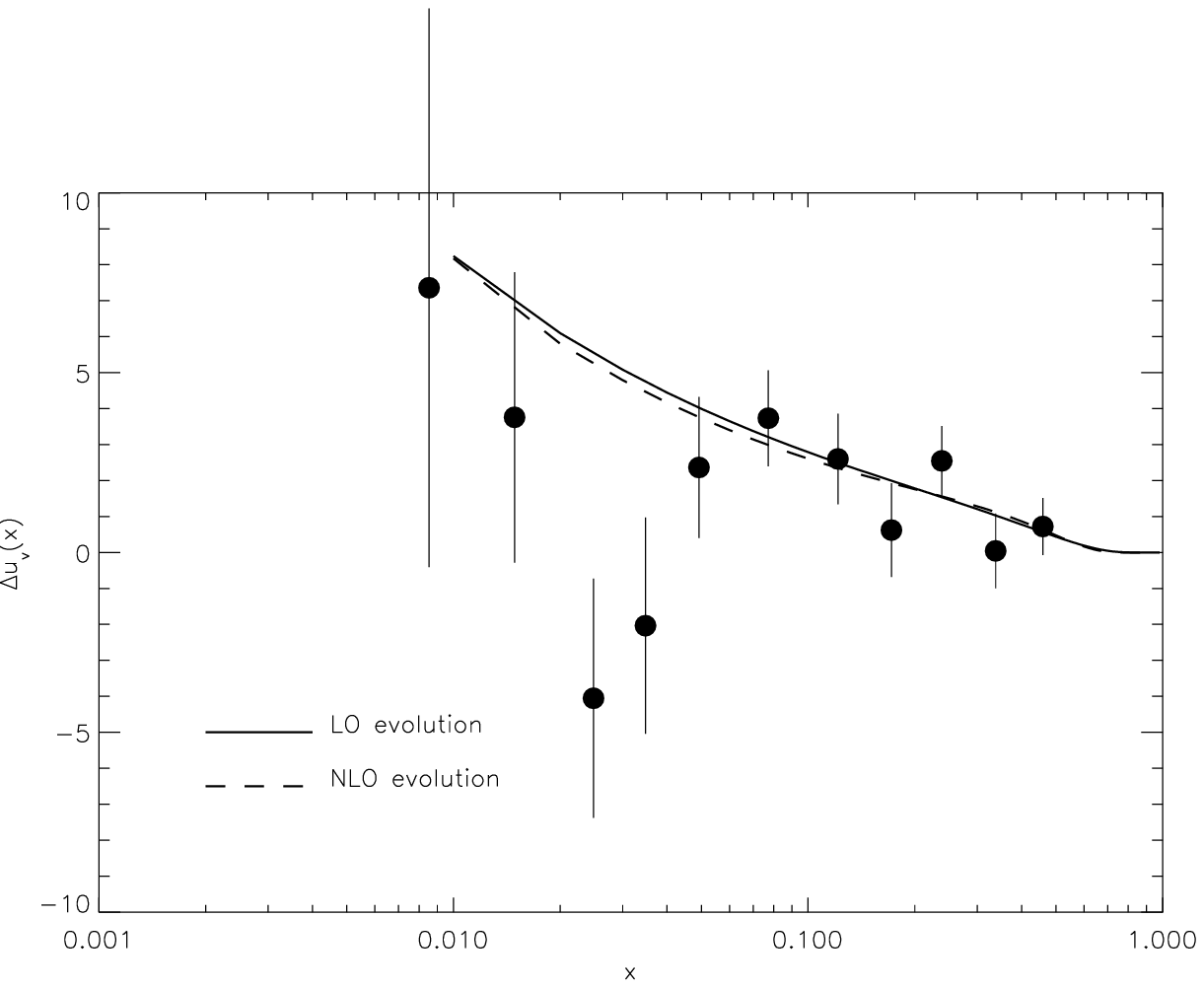}{17}{10}{0}{-10}{1}
\caption{}
\label{fig3}
\end{figure}

\begin{figure}[h]
\vspace{5cm}
\graphics{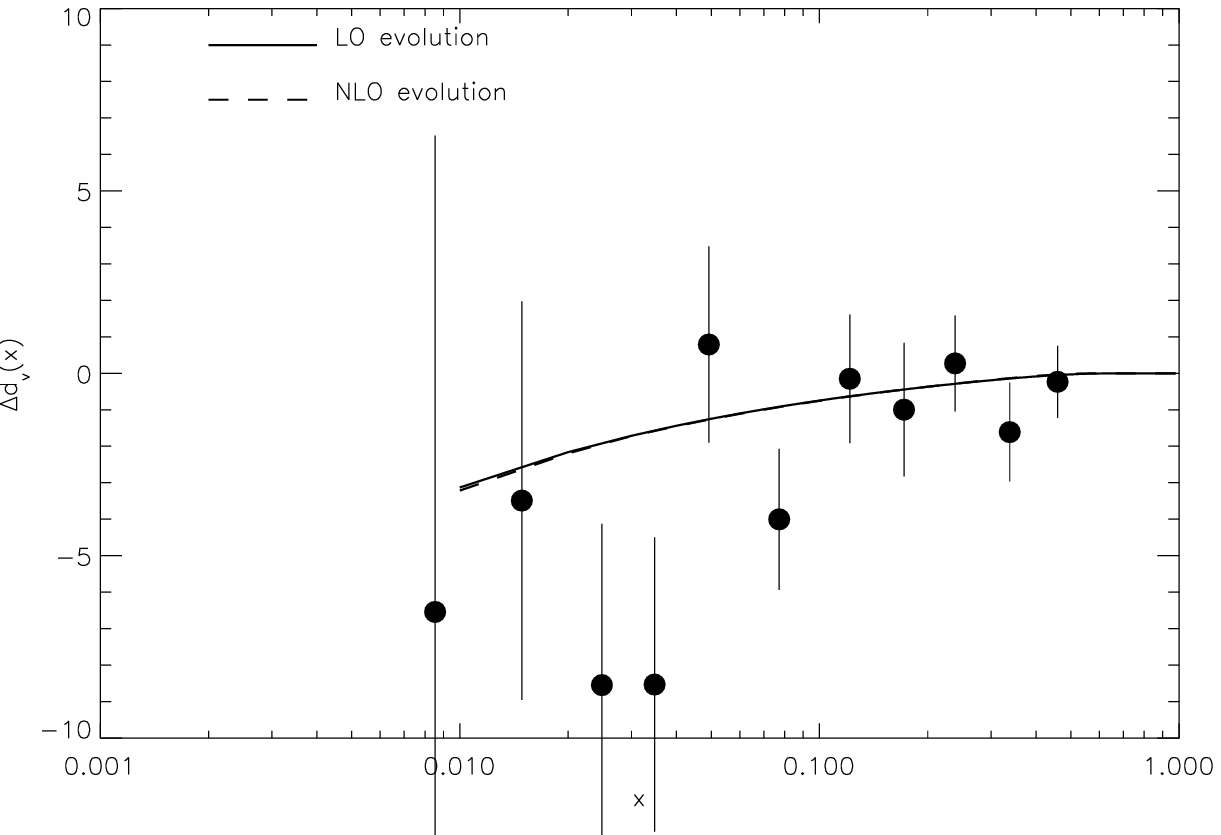}{17}{10}{0}{-10}{1}
\caption{}
\label{fig4}
\end{figure}

Once we have determined the parameters of the model, we are in
the position to calculate other quantities. In Fig. \ref{fig3}
we plot the polarized $u$ valence distribution for the bag model
in LO and NLO against the preliminary experimental data \cite{wislicki}.
There is no real difference between the LO and NLO curves (besides, of
course, the mass parameters). Due to the large errors
in the data it is impossible to conclude anything for the region
$x\le 0.1$. But for $x\ge 0.1$ the theoretical
curves certainly have the correct behaviour. Moreover, the polarized
valence distributions
are pure nonsinglet and we might expect that the bag model would give,
in this case,
a description of data as good as in the unpolarized case. This is
because the problem with the polarized distributions is restricted to the
singlet part of the structure function.
The same considerations can be
drawn from Fig. \ref{fig4}, where the polarized $d$ valence distributions
in LO and NLO are shown together with the data at $Q^2=10\; GeV^2$.

\begin{figure}[h]
\vspace{5cm}
\graphics{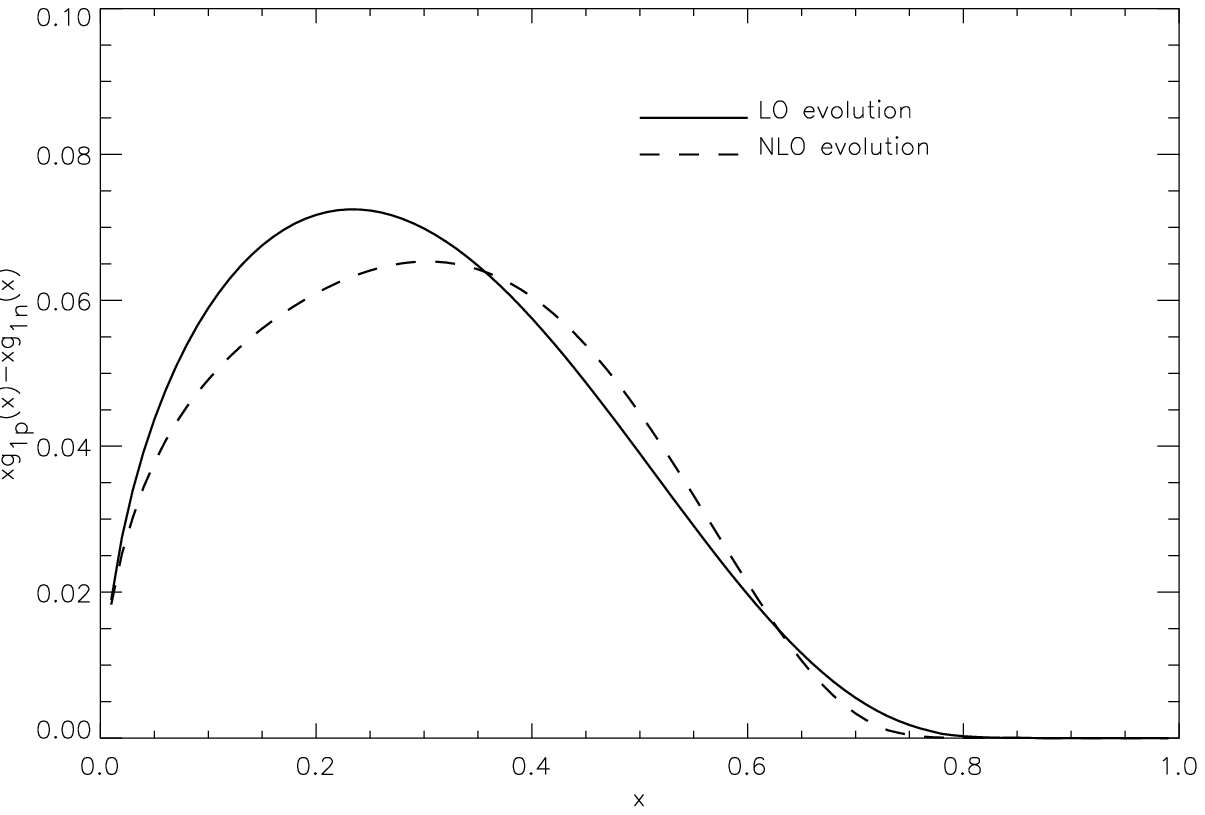}{17}{10}{0}{-10}{1}
\caption{}
\label{fig5}
\end{figure}

Finally, in Fig. \ref{fig5} we show the difference between the polarized
structure functions of the proton and neutron (for which the integral is
$g_A$). Only the theoretical values are presented.According to
(\ref{II8}), in LO the moments of this quantity
are just the moments of the parton distribution $\Delta u(x) - \Delta d(x)$
but in NLO we have to include, in addition to the NLO evolution of the parton
distribution, the second order effects in the form
of the corresponding Wilson coefficient (listed in the Appendix). The Wilson
coefficient leads to a correction in the integral over the NLO curve in Fig.
\ref{fig5}
of the form $1 - \alpha_s (Q^2)/\pi$. In practice, this reduces the integral
from $1.32$ in LO to $1.24$ in NLO ($\alpha_s (Q^2)$ was calculated using
the asymptotic approximation for equ. (\ref{II6}) and it has the value $0.191$
at $Q^2=10 GeV^2$).

\section{Conclusions}

It is clear from the close agreement of the various quark distributions
calculated in LO and NLO that the excellent results obtained in earlier
studies based on the bag model were reliable. That the di-quark masses
required (700 and 900 $MeV$) were so close to the values expected in the
model (600 and 800 $MeV$ - c.f. sect. 4) is very reassuring. Furthermore,
the strong coupling constant in NLO is small enough that one can be quite
confident in the convergence of the calculation.

It is important to be clear what has and has not been achieved. As most quark
models (including the bag) are {\it not} derived from $QCD$ one cannot
calculate
structure functions unambiguously at NLO. There is an inevitable limit
to the accuracy of any quark model. What the present work has shown is that
once the parameters of the model (e.g. $\mu^2$, $R$, $M_s$, $M_v$)
are fixed by the unpolarized valence distribution in either LO and NLO,
the predictions for the spin dependent valence quark distributions
are determined relatively unambiguously. This is an important result as it
does give one some confidence in our ability to investigate problems like
those mentioned in the introduction (nuclear EMC effect, spin crisis, etc.)
within quark models.

As an immediate application we plan to estimate the effect of the pion
cloud of the nucleon on the spin and flavour dependence of the bag
structure functions \cite{nos}.
\newline
\newline
\newline
We thank to A. Schreiber for providing the program for leading order
evolution which we extended to next-to-leading order. F.M.S
wishes to thank to W. Melnitchouk for helpful discussions.
This work was supported by the Australian Research
Council and by CAPES (Brazil).

\appendix
\makeatletter                    
\@addtoreset{equation}{section}  
\makeatother                     
\renewcommand{\theequation}{\thesection.\arabic{equation}}
\section{Appendix: Beta function, NS anomalous dimension and Wilson coeficient}

We here list the coefficients appearing in the expansions
(\ref{II3})--(\ref{II5}). They can be found in refs.
\cite{gross,jones,kodaira}.

\begin{equation}
\beta_0 = \frac{11 C_A -4 T_R N_f}{3},
\end{equation}
\begin{equation}
\beta_1 = \frac{34}{3} C_A^2 -4\left(\frac{5}{3}C_A + C_F\right)T_R N_f ,
\end{equation}

\begin{equation}
\gamma^{(0) n} = 2C_F \left(-3-\frac{2}{n(n+1)} + 4 S_1 (n) \right),
\end{equation}

\begin{eqnarray}
\gamma^{(1) n}&=& (C_{F}^{2}-\frac{1}{2}C_F C_A)\left\{16 S_1 (n)
\frac{2n+1}{n^2 (n+1)^2} + 16 \left[2 S_1 (n) - \frac{1}{n(n+1)}\right]
[S_2(n) - S'_{2}\left(\frac{n}{2}\right)] \right. \nonumber \\*
      && \left. + 64 \tilde S (n) + 24 S_2 (n) - 3 - 8
S'_{3}\left(\frac{n}{2}\right)
- 8\frac{3 n^3 + n^2 -1}{n^3 (n+1)^3} - 16 (-1)^{\eta}
\frac{2 n^2 + 2n + 1}{n^3 (n+1)^3}\right\} \nonumber \\*
      && + C_F C_A \left\{S_1 (n)\left[\frac{536}{9} + 8\frac{2n + 1}
{n^2 (n+1)^2}\right] - 16 S_1 (n) S_2 (n) \right. \nonumber \\*
      && \left. + S_2 (n) \left[-\frac{52}{3} + \frac{8}{n(n+1)}\right]
-\frac{43}{6} -4\frac{151 n^4 +263 n^3 +97 n^2 +3n +9}{9 n^3 (n+1)^3}\right\}
\nonumber \\*
      && + \frac{C_F N_f}{2}\left\{-\frac{160}{9}S_1 (n) + \frac{32}{3}S_2 (n)
+ \frac{4}{3} + 16\frac{11 n^2 + 5n + -3}{9 n^2 (n+1)^2}\right\},
\end{eqnarray}

\begin{eqnarray}
F_n &=& C_F \left(-9 +\frac{1}{n} + \frac{2}{n+1} + \frac{2}{n^2}
+ 3 S_1 (n) -4 S_2 (n) - \frac{2}{n(n+1)}S_1 (n) \right. \nonumber \\*
    && \left.+ 2 (S_1^2 (n) + S_2 (n))\right).
\end{eqnarray}
For the case of $SU(3)_c$, $C_F = \frac{4}{3}$, $C_A = 3$ and $T_R =
\frac{1}{2}$. The parameter $\eta$ is 1 or 2 for analytic continuation of
odd or even momenta respectively.
The functions $S (n)$ are given by:
\begin{equation}
S_i (n) = \sum_{j=1}^{n}\frac{1}{j^i},
\end{equation}
\begin{equation}
S'_{i}\left(\frac{n}{2}\right)=\frac{1+(-1)^{\eta}}{2}S_{i}\left(\frac{n}{2}\right)
+ \frac{1-(-1)^{\eta}}{2}S_{i}\left(\frac{n - 1}{2}\right),
\end{equation}
\begin{equation}
\tilde S (n)=\sum_{j=1}{n}\frac{(-1)^j}{j^2} S_1 (j).
\end{equation}

We use the inverse Laplace transformation of the IMSL library
to make the inversion of expression (\ref{II7}) for the moments
and extract the parton distributions. Once the moments are defined only for
$n$ even or odd, it is necessary to extend the validity of the
equations for the whole interval of $n$. This is done through
analytic continuation in which case the values of the anomalous
dimensions and Wilson coefficients starting from even or odd $n$
coincide. For this purpose, it is better to work
with the forms of the $S(n)$ functions given in
Appendix $A$ of the 1990 paper of Gl\"{u}ck et al. \cite{gluck}.

\addcontentsline{toc}{chapter}{\protect\numberline{}{References}}

\newpage

Figure 1: Total valence distribution in the bag compared with the
MRS \cite{mrs} parametrization of the data in the $\overline{MS}$ scheme.
The quark distributions are evolved in leading order
$QCD$.

Figure 2: Total valence distribution in the bag compared with the
MRS \cite{mrs} parametrization of the data in the $\overline{MS}$ scheme.
The quark distributions are evolved in next-to-leading order $QCD$.

Figure 3: Preliminary polarized valence data for the
up quark distribution in the proton \cite{wislicki} compared with bag
model predictions in leading order and next-to-leading order.

Figure 4: Preliminary polarized valence data for the
down quark distribution in the proton \cite{wislicki} compared with bag
model predictions in leading order and next-to-leading order.

Figure 5: Bag model prediction for the $x$ dependence of the
Bjorken sum rule in leading order and next-to-leading order
$QCD$ evolution.

\end{document}